# Comment: The 2005 Neyman Lecture: Dynamic Indeterminism in Science

**Hans R. Künsch**

## 1. INTRODUCTION

Professor Brillinger is to be congratulated on this paper which is both a contribution to the history of statistics and an introduction to statistical modeling using stochastic processes, a topic that continues to be of great relevance in theory and applications. It is interesting to see how many ideas have been formulated already at an early stage. In particular, I like the idea of "synthetic data" to judge the adequacy of a fitted model. With time series or spatial data, one typically needs only a few replicates to assess visually the differences between real and synthetic data, and so this is really a powerful tool.

## 2. COMMENTS ON THE DATA EXAMPLES

If I understand the description of the data behind Figure 4 correctly, the rainfall has been averaged over 53 seeding days. I would expect the wind speeds to vary from day to day, so I would use a hierarchical model for the wind speeds $v_j$ with a variance component within the same day and a variance component between days. The variation between days would lead to some variation of the time of the peak, and averaging would smear it out. Hence the sharp peak in Figure 4 is even more surprising. The only possibility I see for a model that produces a similar peak as in the actual data, is to assume a decaying intensity for the process of rain particles in Ticino.

In the two population dynamics examples, no full probabilistic model is constructed. Only the conditional mean values and not the distribution of the fluctuations are considered. Least squares methods are used for fitting. Moreover, the reproduction process is not part of the model, although the reader is referred to Guttorp (1980) for a treatment of births in the second example. From a pragmatic point, it can certainly be advantageous to focus on those parts that are of primary concern without making assumptions on other processes in the system. If only the population above a certain threshold age are of interest and if information about the number of individuals reaching the threshold age is available, then one does not need to model the births. On the other hand, as I will argue below, there is also the point of view that in order to understand a system, all relevant processes should be included.

As an interesting complement to Example 7, I would like to mention the paper Jonsen, Mills Flemming and Myers (2005) which also analyses seal movement data. They use a discrete-time integrated random walk for the animal movements, with interpolation to accommodate irregular observation times, and $t$-distributions for the observation errors. With such a model, they can use state-space methodology to fit the model to the data, without having to exclude suspicious observations. Including a drift component to the integrated random walk is possible, but would make the analysis more complicated.

## 3. DETERMINISM AND INDETERMINISM

Even 50 years after Neyman's work discussed in this paper, many fields of science are still dominated by deterministic models, at least in the area of environmental modeling where I have most experience. The reasons for this dominance are that scientists are interested in models that

- take as much knowledge about the underlying processes into account as possible,
- contribute to the understanding of these processes,
- are transferable to similar systems,
- allow prediction of the same system under different driving conditions than those observed,
- have parameters with a clear subject matter interpretation.

*Professor, Seminar für Statistik, Department of Mathematics, ETH Zurich, CH-8092 Zurich, Switzerland (e-mail: kuensch@stat.math.ethz.ch).*

Some of these reservations can be made against the analysis of the weather modification experiment described in the paper. No attempt is made to connect the data with physical knowledge about atmospheric processes in the alpine region on experimental days, and even if the model gave a satisfactory fit it would not be clear how the estimated velocity distribution could be transfered to a slightly different location.

It is more difficult than one might think to include all relevant knowledge about the underlying processes into simple statistical models. One reason is that this knowledge can be described mathematically only in continuous time and space, using ordinary or partial differential equations, whereas statistical models tend to be in discrete time and space. Another reason is that physical models typically involve many variables that are not observed, and this complicates the statistical analysis.

It is clear, however, that even the best available deterministic models are limited because initial conditions, boundary conditions or inputs are uncertain and because the computational complexity of such models requires essential simplifications to make them tractable. As a result, when fitting ordinary and partial differential equation models to data by nonlinear least squares, one often finds that there are systematic deviations between model outputs and observations that cannot be explained by measurement errors alone. Hence, in my view, the challenge for statisticians is to develop methods which build upon the deterministic models in science and at the same time allow to describe uncertainties in realistic ways or even to enhance understanding or improve model extrapolation.

I concur very much with Professor Brillinger that statisticians should make more use of SDEs in their modeling. They are convenient not only because they can handle observations at irregular time points, but rather because they allow to introduce uncertainty into ODEs. However, one should be aware that the noise term has a profound impact on the behavior of the solutions: They are no longer smooth, but have infinite variation on any small interval. This is definitely not realistic for the tracks of elks and seals. In fact most natural systems are believed to evolve smoothly, at least on the macroscopic scale. This makes it difficult to decide which statistical methods one should use when fitting SDE models to data since we do not really believe in the fine structure of the model.

Adding a white noise disturbance to an ODE is not the only way to introduce uncertainty into deterministic modeling approaches. Kennedy and O'Hagan (2001) and Bayarri et al. (2007) introduce—in addition to the measurement error term—a model inadequacy, or bias, term that is intended to capture the effect of model deficiencies on model output. It is usually formulated in a nonparametric form as a Gaussian process with a suitable covariance function. While this approach is universally applicable and can lead to more reliable uncertainty estimates, its use for diagnosing possible causes for model deficits is limited. An alternative consists in replacing a constant model parameter or a deterministic input by a stochastic process model, in the spirit of the quote from Neyman by David Brillinger in Section 3.3 "...with coefficients that are not constants, but random variables." Such a time-varying parameter or input can be estimated jointly with the other (time-constant) parameters, and from a careful analysis of the estimated trajectory additional insight into the nature of model deficits can be gained. This approach has been used in the systems analysis literature for more than 20 years; see, for example, Beck (1983) or Kristensen, Madsen and Jørgensen (2004). These authors use a discrete-time setting and extended Kalman filter techniques. Tomassini et al. (2007) develop a MCMC algorithm that can be used in continuous time and without any linearization technique.

To illustrate the differences between adding a noise term to a differential equation and making a parameter time-varying, consider the simple growth model

$$dx_t = \beta x_t \, dt.$$

The solution of the corresponding SDE

$$dX_t = \beta X_t \, dt + \sigma X_t \, dB_t$$

is $X_t = X_0 \exp((\beta - \sigma^2/2)t + \sigma B_t)$. Hence not only the local, but also the long-time behavior is changed by the added noise. This SDE can be interpreted as saying that we add to the growth rate $\beta$ of the deterministic model a white noise term. This is hardly plausible biologically, and I thus prefer the version

$$d\beta_t = -\gamma(\beta_t - \bar{\beta}) \, dt + \sigma \, dB_t,$$
$$dX_t = \beta_t X_t \, dt.$$

A mean-reverting Ornstein–Uhlenbeck process has been chosen for the dynamics of $\beta_t$ because of its simplicity. It has the drawback of allowing negative values, but other choices are possible.

## 4. TECHNICAL DIFFICULTIES WITH STOCHASTIC DIFFERENTIAL EQUATIONS

I have discussed reasons to use SDE models in statistics. David Brillinger has shown some simple and pragmatic approaches for fitting and analyzing them, mainly by relying on the Euler (or Euler–Maruyama) approximation (3). If one wants to refine this approach and consider, for instance, exact maximum likelihood estimation, then a number of technical difficulties occur. I would like to give a brief overview of these difficulties and modern approaches to solve them since this is a very active area of research at the moment.

### 4.1 Exact Observations at Discrete Time Points

Let us consider the SDE

$$dX_t = \mu(X_t, \theta)\,dt + \sigma(X_t, \theta)\,dB_t$$

with known initial condition $x_0$ and unknown parameter $\theta$. If we observe the solution $(X_t)$ at discrete time points $t_i$, then the log likelihood can be written as

$$\sum_i \log p_\theta(t_{i+1} - t_i, x_{t_i}, x_{t_{i+1}}).$$

The transition densities $p_\theta(t, x, y)$ are, however, not available in closed form, and the Euler approximation implied by (3) is often not accurate enough (the resulting estimator is usually not consistent if the distances $t_{i+1} - t_i$ remain fixed). Numerical computation is difficult because one has to solve a partial differential equation (Fokker–Planck). This can be done in one dimension (see Lindström (2007)), but in higher dimensions it is not practical. Estimating equations other than the MLE have been considered, but for them one also has to compute conditional expectations of the form

$$E_\theta[\psi(X_s, X_t, \theta) \mid X_s]$$

for $s < t$ which often cannot be done in closed form.

The emphasis in much of the recent work has been on Monte Carlo methods. In the framework of estimating equations, this has been developed by Kessler and Paredes (2002). The nice feature of their approach is that if one estimates the conditional expectation by $J$ replicates, the asymptotic variance of the corresponding estimator increases by a factor $(1 + 1/J)$. This means that a small number of replicates is sufficient for all practical purposes.

Monte Carlo methods can also be used for likelihood inference. A natural approach is to consider the values of the process $(X_t)$ on a fine grid between observation times as latent variables and to use the Euler approximation for these smaller time steps. The latent values can then be integrated out using importance sampling, or one can apply the EM-algorithm. Since the E-step rarely can be done analytically, one has to use a Monte Carlo method instead. For integrating the latent variables out, Durham and Gallant (2002) have proposed clever importance distributions which are crucial for the method to become useful. As discussed by Roberts and Stramer (2001), the EM-algorithm suffers from poor convergence if the diffusion coefficient $\sigma$ depends on unknown parameters. The reason for this is that the precision for estimating the diffusion coefficient $\sigma$ of an SDE goes to infinity as the observation points become dense in an interval, or in other words, the distributions of the solution of two SDEs with different $\sigma$ are mutually singular. To overcome this problem, Roberts and Stramer (2001) propose a transformation of the latent variables that reduces the information they contain about $\sigma$. The same problem affects also the first approach: One cannot use the same importance distribution for parameter values that correspond to different values of $\sigma$. Hence one cannot estimate the whole likelihood function by a single simulation experiment.

An entirely different approach has been used by Beskos, Papaspiliopoulos, Roberts and Fearnhead (2006). Based on exact simulation methods for diffusions, they propose several unbiased estimators of the likelihood function and a stochastic EM-algorithm. The ideas in this paper are most interesting, but unfortunately they cannot be used for arbitrary SDEs. It must be possible to transform the variables so that the diffusion coefficient is constant, and the drift coefficient $b$ must be derived from a potential function.

### 4.2 Partial and Noisy Observations

In many cases, we do not observe $X_{t_i}$ exactly, but only random variables $Y_i$ which are conditionally independent and such that $Y_i$ depends on $X_{t_i}$ only:

$$Y_i \mid X_{t_i} \sim f(y \mid x_{t_i})\,dy.$$

We are then in the setup of state-space models. Particle filter methods (see, e.g., Doucet, de Freitas and Gordon (2001)) can be used to estimate both the likelihood function and the unobserved path $(X_t)$. Again, the unavailability of the transition densities creates additional problems. One can either extend the state variables by adding the values of $(X_t)$ on a fine grid between observation times, or one can use again ideas from exact sampling of diffusions; see Fearnhead, Papaspiliopoulos and Roberts (2008). The issue of efficient computations for estimating parameters remains an open problem.

Ramsay, Hooker, Campbell and Cao (2007) have considered an approach to estimate simultaneously an approximate solution of an ODE together with unknown parameters $\theta$ by minimizing

$$-\sum_i \log f(y_i \mid x_{t_i}) + \lambda \int_0^T \left\| \frac{d}{dt} x_t - \mu(x_t, \theta) \right\|^2 dt.$$

Numerically, this minimization problem is solved by approximating $(x_t)$ in a spline basis. The approach can be viewed as MAP estimation in an SDE model with drift $\mu$ and constant diffusion coefficient $1/\sqrt{2\lambda}$. In the case of a more general diffusion coefficient, one can replace the second term by

$$\int_0^T \left\| \sigma(x_t, \theta)^{-1} \left( \frac{d}{dt} x_t - \mu(x_t, \theta) \right) \right\|^2 dt.$$

I think that for estimating parameters, this approach is potentially simpler than those based on the particle filter.